\documentclass[prl,twocolumn,amsmath,amssymb,superscriptaddress,
showpacs]{revtex4}
\usepackage{graphics}
\usepackage{epsfig}
\usepackage{bbm}
\usepackage[latin1]{inputen c} 

\newcommand{\be}{\begin{equation}}
\newcommand{\ee}{\end{equation}}
\newcommand{\eea}{\end{eqnarray}}

\newcommand{\va}[1]{\ensuremath{(\Delta#1)^2}}

\newcommand{\exs}[1]{\ensuremath{\langle{#1}\rangle}}

\newcommand{\qed}{\ensuremath{\hfill \Box}}

\newcommand{\ketbra}[1]{\ensuremath{| #1 \rangle \langle #1 |}}

\newcommand{\ket}[1]{\ensuremath{|#1\rangle}}

\newcommand{\kommentar}[1]{}

\begin{document}
\title{
Optimal spin squeezing inequalities detect bound entanglement in
spin models}
\date{\today}
\begin{abstract}
We determine the complete set of generalized spin squeezing
inequalities. These are entanglement criteria that can be used for
the experimental detection of entanglement in a system of
spin-$\tfrac{1}{2}$ particles in which the spins cannot be
individually addressed. They can also be used to show the presence
of bound entanglement in the thermal states of several spin models.
\end{abstract}

\author{G\'eza T\'oth}
\affiliation{ICFO-Institut de Ci\`encies Fot\`oniques, E-08860
Castelldefels (Barcelona), Spain} \affiliation{Research Institute
for Solid State Physics and Optics, Hungarian Academy of Sciences,
 P.O. Box 49, H-1525 Budapest,
Hungary}

\author{Christian Knapp}
\affiliation{Institut f\"ur Theoretische Physik, Universit\"at
Innsbruck, Technikerstra{\ss}e 25, A-6020 Innsbruck, Austria,}

\author{Otfried G\"uhne}
\affiliation{Institut f\"ur Quantenoptik und Quanteninformation,
\"Osterreichische Akademie der Wissenschaften, A-6020 Innsbruck,
Austria}

\author{Hans J. Briegel}

\affiliation{Institut f\"ur Theoretische Physik, Universit\"at
Innsbruck, Technikerstra{\ss}e 25, A-6020 Innsbruck, Austria,}
\affiliation{Institut f\"ur Quantenoptik und Quanteninformation,
\"Osterreichische Akademie der Wissenschaften, A-6020 Innsbruck,
Austria}


\maketitle



Entanglement lies at the heart of many problems in quantum mechanics
and has attracted increasing attention in recent years. However,
in spite of intensive research, many of its intriguing properties
are not fully understood. For example, it has been shown that there
are entangled states, from which the entanglement cannot be
distilled again into the pure state form, even if many copies of the
state are available \cite{bound}. The existence of these so-called
bound entangled states has wide-ranging consequences for quantum
cryptography \cite{boundqkd} and classical information theory
\cite{boundinformation}. Since entangled states that are not
recognized by the separability criterion of the positivity of the
partial transpose (PPT) \cite{ppt} are bound entangled, such states
also serve as a test bed for new separability criteria \cite{ccn,
doherty, ghge}. However, bound entangled states are often considered
to be rare, in the sense that they do not occur under natural
conditions.

In physical systems such as ensembles of cold atoms \cite{HS99} or
trapped ions \cite{VR01}, spin squeezing \cite{K93,W94} is one of
the most successful approaches for creating large scale quantum
entanglement. Since the variance of a spin component is small, spin
squeezed states can be used for reducing spectroscopic noise or to
improve the accuracy of atomic clocks \cite{K93,W94}. Moreover, if
an $N$-qubit state violates the inequality \cite{SD01}
\begin{eqnarray}
\frac{\va{J_x}}{\exs{J_y}^2+\exs{J_z}^2}\ge \frac{1}{N},
\label{motherofallspinsqueezinginequalities}
\end{eqnarray}
where $J_l:=\tfrac{1}{2}\sum_{k=1}^N \sigma_{l}^{(k)}$ for $l=x,y,z$
are the collective angular momentum components and
$\sigma_{l}^{(k)}$ are Pauli matrices, then the state is entangled
(i.e., not separable), which is necessary for using it in quantum
information processing applications \cite{SD01}.

Recently, several generalized spin squeezing criteria for the
detection of entanglement appeared in the literature
\cite{GT04,KC05,GT06} and have been used experimentally
\cite{spexp}. These criteria have a large practical importance since
in many quantum control experiments the spins cannot be individually
addressed, and only collective quantities can be measured. In
Ref.~\cite{KC05} a generalized spin squeezing criterion was
presented detecting the presence of two-qubit entanglement. In Refs.
\cite{GT04,GT06} other criteria can be found that detect
entanglement close to spin singlets and symmetric Dicke states,
respectively. These entanglement conditions were obtained using very
different approaches. At this point two main questions arise: (i) Is
there a systematic way of finding all such inequalities? Clearly,
finding such optimal entanglement conditions is a hard task since
one can expect that they contain complicated nonlinearities. (ii)
How strong are spin squeezing criteria? Can they detect entangled
states that are not detected by the PPT criterion or other known
entanglement criteria?

\begin{figure}
\centerline{ \epsfxsize3.3in \epsffile{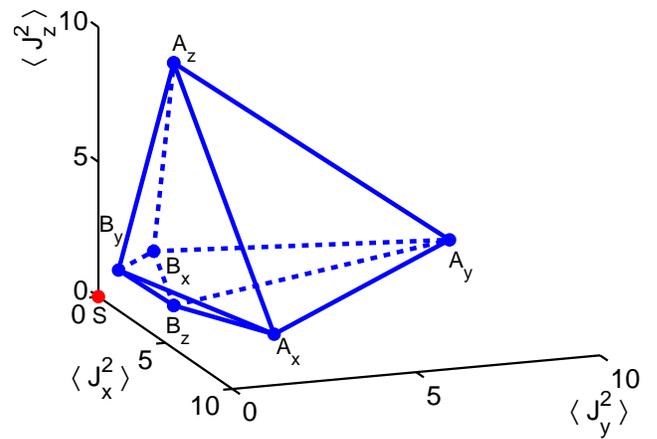}} \caption{The
polytope of separable states corresponding to Eqs.~(\ref{Jxyzineq})
for $N=6$ and $\vec{J}=0.$ $S$ corresponds to a many body singlet
state.} \label{J2xyz}
\end{figure}

The goal of this Letter is twofold. First, we give  a complete set
of spin squeezing inequalities based on the first and second moments
of collective observables. Second, we use them to show the presence
of multipartite bound entanglement in several spin models in thermal
equilibrium. In particular, we consider bound entanglement that has
a positive partial transpose with respect to all bipartitions.

We can directly formulate our first main result:
\\
{\bf Observation 1.} {\it Let us assume that for a physical system
the values of $\vec{J}:=(\exs{J_x},\exs{J_y},\exs{J_z})$ and
$\vec{K}:=(\exs{J_x^2},\exs{J_y^2},\exs{J_z^2})$ are known.
Violation of any of the following inequalities implies entanglement:
\begin{subequations}
\begin{eqnarray}
\exs{J_x^2}+\exs{J_y^2}+\exs{J_z^2} &\le& N(N+2)/4,
\label{theorem1a}
\\
\va{J_x}+\va{J_y}+\va{J_z} &\ge& N/2, \label{Jxyzineq_singlet}
\\
\exs{J_i^2}+\exs{J_j^2}-N/2 &\le& (N-1)\va{J_k},
\label{Jxyzineq_spsq2}
\\
(N-1)\left[\va{J_i}+\va{J_j}\right] &\ge& \exs{J_k^2}+N(N-2)/4,
\label{Jxyzineq_spsq3} \;\;\;\;\;\;
\end{eqnarray}
\label{Jxyzineq}
\end{subequations}
where $i,j,k$ take all the possible permutations of $x,y,z.$ The
proof is given in the Appendix. }

For any value of $\vec{J}$ these eight inequalities define a
polytope in the three-dimensional
$(\exs{J_{x}^2},\exs{J_{y}^2},\exs{J_{z}^2})$-space. Observation 1
states that separable states lie inside this polytope. For the case
$\vec{J}=0$ and $N=6$ the polytope is depicted in Fig.~\ref{J2xyz}.
Such a polytope is completely characterized by its extreme points.
Direct calculation shows that they are
\begin{align}
A_x &:=\left[ \frac{N^2}{4}-\kappa(\exs{J_y}^2+\exs{J_z}^2),
\frac{N}{4}+\kappa\exs{J_y}^2, \frac{N}{4}+\kappa\exs{J_z}^2
\right], \nonumber
\\
B_x&:=\left[ \exs{J_x}^2+\frac{\exs{J_y}^2+\exs{J_z}^2}{N},
\frac{N}{4}+\kappa \exs{J_y}^2, \frac{N}{4}+\kappa\exs{J_z}^2
\right], \nonumber
\end{align}
where $\kappa:=(N-1)/N.$ The points $A_{y/z}$ and $B_{y/z}$ can be
obtained in an analogous way.

One might ask whether all points inside the polytope correspond to
separable states. This would imply that the criteria of Observation
1 are complete, that is, if the inequalities are satisfied, then the
first and second moments of $J_k$ do not suffice to prove
entanglement. In other words, it is not possible to find criteria
detecting more entangled states based on these moments. Due to the
convexity of the set of separable states, it is enough to
investigate the extreme points: 
\\
{\bf Observation 2.} {\it For any value of $\vec{J}$ there are
separable states corresponding to $A_k.$ For certain values of
$\vec{J}$ and $N$ there are separable states corresponding to points
$B_k.$ However, there are always separable states corresponding to
points $B_k'$ such that their distance from $B_k$ is smaller than
$1/4.$ In the limit $N\rightarrow \infty$ for a fixed normalized
angular momentum $\vec{j}:=\vec{J}/(N/2)$ the difference between the
volume of polytope of Eqs.~(\ref{Jxyzineq}) and the volume of set of
points corresponding to separable states decreases with $N$ at least
as $\Delta V/V \propto N^{-2},$ hence in the macroscopic limit the
characterization is complete. }

{\it Proof.} A separable state corresponding to $A_x$ is
\begin{equation}
\rho_{A_x}:=p(\ketbra{\psi_+})^{\otimes N}+
(1-p)(\ketbra{\psi_-})^{\otimes N}. \label{Ax}
\end{equation}
Here $\ket{\psi_{+/-}}$ are the single qubit states with Bloch
vector coordinates $(\exs{\sigma_x},\exs{\sigma_y},\exs{\sigma_z})=
(\pm c_x,\exs{J_y}/J,\exs{J_z}/J)$ where $J:=N/2$ and
$c_x:=\sqrt{1-(\exs{J_y}^2+\exs{J_z}^2)/J^2}.$ The mixing ratio is
defined as $p:=[1+\exs{J_x}/(J c_x)]/2.$ If $M:=Np$ is an integer,
we can also define the state corresponding to the point $B_x$ as \be
\ket{\phi_{B_x}}:=\ket{\psi_+}^{\otimes M}\otimes
\ket{\psi_-}^{\otimes (N-M)}.\ee If $M$ is not an integer, we can
approximate $B_x$ by taking $m:= M-\varepsilon$ as the largest
integer smaller than $M,$ defining $\rho':=(1-\varepsilon)
(\ketbra{\psi_+})^{\otimes m} \otimes
(\ketbra{\psi_-})^{\otimes(N-m)} + \varepsilon
(\ketbra{\psi_+})^{\otimes (m+1)} \otimes (\ketbra{\psi_-})^{\otimes
(N-m-1)}.$ This state has the same coordinates as $B_x,$ except for
the value of $\exs{J_x^2},$ where the difference is $ c_x^2
(\varepsilon- \varepsilon^2) \le 1/4.$ The dependence of $\Delta
V/V$ on $N$ can be studied by considering the polytopes in the
$(\exs{J_{x}^2},\exs{J_{y}^2},\exs{J_{z}^2})$-space corresponding to
$\exs{J_k}=j_k\times N/2,$ where $j_k$ are the normalized angular
momentum cordinates. As $N$ increases, the distance of the points
$A_k$ to $B_k$ scales as $N^2,$ hence the volume of the polytope
increases as $N^6.$ The difference between the polytope and the
points corresponding to separable states scales like the surface of
the polytope, hence as $N^4.$ \qed

Now we consider already known entanglement criteria and show how
they  can be derived from our theory. This can be done by showing
that for any $\vec{J}$ the  points $A_k$ and $B_k$ satisfy them.

{\it Case 1.} The standard spin-squeezing inequality is
Eq.~(\ref{motherofallspinsqueezinginequalities}) from
Ref.~\cite{SD01}. This inequality is valid for all $A_k$ and $B_k,$
for $B_x$ even equality holds.

{\it Case 2}. For separable states $\exs{J_x^2}+\exs{J_y^2} \le
(N^2+N)/4$ holds \cite{GT06}, as can be proved in the same way. This
can be used to detect entanglement close to the $N$-qubit symmetric
Dicke states with $N/2$ excitations.

{\it Case 3}. Separable states fulfill Eq.~(\ref{Jxyzineq_singlet})
which has already been shown in Ref.~\cite{GT04}. It is maximally
violated by a many-body singlet, e.g., the ground state of an
anti-ferromagnetic Heisenberg chain.

{\it Case 4}. For symmetric states it is known that
$\exs{J_x^2}+\exs{J_y^2}+\exs{J_z^2}=N(N+2)/4$ \cite{KC05}. From
this and Eq.~(\ref{Jxyzineq_spsq2}) one can directly derive
$1-4\exs{J_i}^2/N^2 \leq 4 \va{J_i}/N$ from Ref.~\cite{KC05}.

Next, it is interesting to ask what kind of entanglement is detected
by our criteria knowing that they contain only two-body correlation
terms of the from $\exs{\sigma_k^{(i)} \sigma_k^{(j)}}$ and do not
depend on higher order correlations. In fact, all quantities in our
inequalities can be evaluated based on knowing the average two-qubit
density matrix $\rho_{\rm av2}:=\frac{1}{N(N-1)}\sum_{i\ne
j}\rho_{ij}.$ Do our criteria simply detect entanglement of the
two-qubit reduced state of the density matrix? We will now show that
the criteria Eqs.~(\ref{Jxyzineq}) can detect entangled states that
have a separable two-qubit density matrix. Even more surprisingly,
they can detect bound entanglement in spin systems. While in the
following we will use Eqs.~(\ref{Jxyzineq}) for the theoretical
analysis of spin models, we stress that Eqs.~(\ref{Jxyzineq}) can
also be used for the experimental detection of entanglement in a
realization of these models in physical systems in which the
collective angular momentum can be measured (e.g.,
Ref.~\cite{spexp}).

\begin{figure}
\centerline{ \epsfxsize3.3in \epsffile{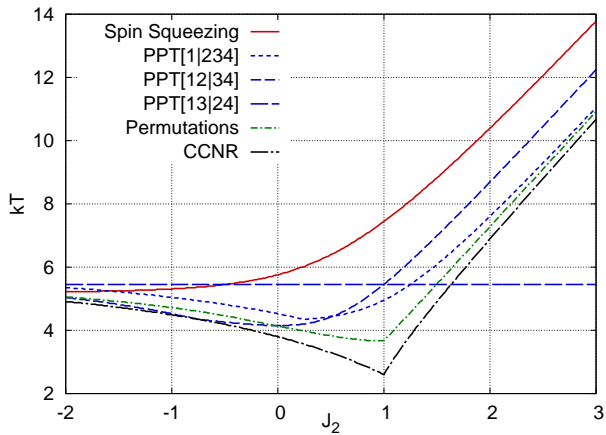}}
\caption{Entanglement properties of the spin model with the
Hamiltonian Eq.~(\ref{cluster}). The critical temperatures for
several entanglement conditions are shown as a function of the
next-to-nearest neighbor coupling $J_2.$ For details see text.}
\label{figcrit}
\end{figure}

Let us first consider four spin-1/2 particles, interacting via the
Heisenberg-type Hamiltonian \cite{tribedi} \be H= \sum_{k=1}^4
\vec{\sigma}_k\vec{\sigma}_{k+1} + J_2 (
\vec{\sigma}_1\vec{\sigma}_3+\vec{\sigma}_2\vec{\sigma}_4).
\label{cluster} \ee where
$\vec{\sigma}=(\sigma_x,\sigma_y,\sigma_z).$ For the above
Hamiltonian, we compute the thermal state $\varrho(T,J_2)\propto
\exp(-H/kT)$ and investigate its separability properties.
Hamiltonians of the type Eq.~(\ref{cluster}) are by no means
artificial: They are used to describe cuprate and polyoxovanadate
clusters \cite{tribedi, chemie}. For several separability criteria
we calculate the maximal temperature, below which the criteria find
the states entangled. The results are summarized in
Fig.~\ref{figcrit}. For $J_2 \gtrsim -0.5$, the spin squeezing
inequality Eq.~(\ref{Jxyzineq_singlet}) is the strongest criterion
for separability. It allows to prove the presence of entanglement
even if the state is PPT with respect to all bipartitions
\cite{ppt}. This implies that the state is multipartite bound
entangled: No pure entangled state can be distilled from it
\cite{duer}. Note that introducing the next-to-nearest neighbor
coupling made the PPT entangled temperature range larger.

For comparison,  we investigated the computable cross norm or
realignment criteria (CCNR, \cite{ccn}) corresponding to all
bipartitions, all the other criteria based on permutations
\cite{clarisse}, and the criterion based on covariance matrices
\cite{ghge}. None of them is able to find PPT entanglement in our
spin system. Finally, we studied for each bipartition the
separability test of symmetric extensions \cite{doherty2} that is
strictly stronger than the PPT criterion. The critical temperatures,
however, coincide within numerical accuracy with the ones from the
PPT criterion, giving strong evidence that $\varrho$ is indeed
separable for the bipartitions. Indeed, we will see later that in
some spin models, the spin squeezing inequalities signal the
presence of entanglement even for states that are separable with
respect to all bipartitions.

\begin{table}
\centerline{ \begin{tabular}{|l l||c|c|c|c|c|c|c|}
  \hline
  &N & 3 & 4 & 5 & 6 & 7 & 8 & 9 \\
  \hline
  Heisenberg &\vline \;\;Eqs.~(\ref{Jxyzineq}) & 5.46 & 5.77 & 5.72 & 5.73 & 5.73 & 5.73 & 5.73 \\
  model &\vline \;\;PPT & 4.33 & 5.47 & 4.96 & 5.40 & 5.17 & 5.37 & 5.25 \\
  \hline
 XY  &\vline\;\;Eqs.~(\ref{Jxyzineq}) & 3.08 & 3.48 & 3.39 & 3.41 & 3.41 & 3.41 & 3.41 \\
 model &\vline\;\;PPT & 2.56 & 3.46 & 3.08 & 3.34 & 3.19 & 3.32 & 3.24
   \\
  \hline
\end{tabular}} \caption{Critical temperatures for the PPT criterion  and Eqs.~(\ref{Jxyzineq})
for Heisenberg and XY spin chains of various size. } \label{tb}
\end{table}

After small spin clusters, we  consider larger spin systems. Using
Eqs.~(\ref{Jxyzineq}), we find bound entanglement that is PPT with
respect to all bipartitions in Heisenberg and XY chains with a
periodic boundary condition with up to 9 qubits.  The critical
temperatures are shown in Table I. Eqs.~(\ref{Jxyzineq}) also
detect bound entanglement in Heisenberg and XY models with a
complete graph topology \cite{GT05}. Latter is a special case of the
Lipkin-Meshkov-Glick model \cite{LMG}. In all these cases there is a
considerable temperature range for which the thermal state is PPT
with respect to all partitions but not yet separable \cite{knapp}.
Interestingly, since in the three-qubit Heisenberg model the thermal
state is invariant under multilateral unitary transformations of the
type $U\otimes U\otimes U,$ for such states the PPT condition
implies biseparability \cite{N3}. Thus, the spin-squeezing
inequalities can detect bound entanglement for which all
bipartitions are separable.

Note that the bound entanglement that is PPT with respect to all
bipartitions is perhaps the most intriguing type and the most
challenging to detect. No pure state entanglement can be distilled
from it with local operations and classical communication, even if
arbitrary number of parties join. However, an entangled state that
is PPT with respect to only a single partition is already bound
entangled since no GHZ state can be distilled from it \cite{duer}.
Such entanglement can be found by the PPT criterion with respect to
a different partition. It is expected to appear in many systems
since as the temperature increases, not all the bipartitions become
PPT at the same temperature \cite{NPTbound}.

Our study of the spin models has two general consequences. First, we
realize that examination of spin models via the partial
transposition or the investigation of bipartitions does not lead to
a full understanding of the entanglement properties of condensed
matter systems. Second, we note that the spin clusters and spin
chains we studied are models of existing physical systems. Thus
multipartite bound entanglement that is PPT with respect to all
partitions is not a rare phenomenon in nature.

Moreover, based on Ref.~\cite{njp}, it is possible to connect the
variances of collective angular momenta to important thermodynamical
quantities giving our inequalities a new physical interpretation.
Let us consider a system with a Hamiltonian $H$ and an additional
magnetic interaction $H_{I} := \sum_{k=x,y,z} B_k J_k,$ where
$\vec{B}$ is the magnetic field. Moreover, assume that
 $H$ commutes with $J_{x/y/z}.$ Then
the magnetic susceptibilities are $\chi_l := (\partial \exs{J_l} /
\partial B_l)\vert_{\vec{B}=0}$ for $l=x,y,z$ and the variances can be
written as
 $\va{J_l}=kT\chi_l.$ Thus our inequalities can be expressed
 with susceptibilities \cite{remark}.

Finally, we discuss some further features of our spin squeezing
inequalities. One can ask what happens, if not only $\exs{J_k}$ and
$\exs{J^2_k}$ for $k=x,y,z$ are known, but $\exs{J_{i}}$ and
$\exs{J^2_{i}}$ in arbitrary directions $i$. We will now show how to
find the optimal directions $x',y',z'$ to evaluate Observation 1.
Knowledge of $\exs{J_{i}}$ and $\exs{J^2_{i}}$ in arbitrary
directions is equivalent to the knowledge of the vector $\vec{J},$
the correlation matrix $C$ and the covariance matrix $\gamma,$
defined as
\cite{ghge, DU06}
$C_{kl}:={\exs{J_k J_l+J_l J_k}}/{2}$ and
$\gamma_{kl}:=C_{kl} - \exs{J_k}\exs{J_l}$
for $k,l=x,y,z.$ When changing the coordinate system to $x',y',z',$
vector $\vec{J}$ and the matrices $C$ and $\gamma$ transform as
$\vec{J} \mapsto O \vec{J},$ $C\mapsto O C O^T$ and $\gamma \mapsto
O \gamma O^T$ where $O$ is an orthogonal $3\times3$-matrix. Looking
at the inequalities of Observation 1 one finds that the first two
inequalities are invariant under a change of the coordinate system.
Concerning Eq.~(\ref{Jxyzineq_spsq2}), we can reformulate it as
$\exs{J^2_i}+\exs{J^2_j}+\exs{J^2_k}-N/2 \leq (N-1) \va{J_k} +
\exs{J^2_k}.$ Then, the left hand side is again invariant under
rotations, and we find a violation of Eq.~(\ref{Jxyzineq_spsq2}) in
some direction if the minimal eigenvalue of
$\mathfrak{X}:=(N-1)\gamma+C$ is smaller than $Tr(C)-N/2.$
Similarly, we find a violation of Eq.~(\ref{Jxyzineq_spsq3}) if the
largest eigenvalue of $\mathfrak{X}$ exceeds
$(N-1)Tr(\gamma)-N(N-2)/4.$ Thus the orthogonal transformation that
diagonalizes $\mathfrak{X}$ delivers the optimal measurement
directions $x',y',z'$ \cite{qubit4matlab}.

In summary, we presented a family of entanglement criteria that
detect any entangled state that can be detected based on the first
and second moments of collective angular momenta. We applied our
findings to examples of spin models, showing the presence of bound
entanglement in these models.

 We thank A. Ac\'{\i}n, J.I. Cirac, J.
Korbicz, and M. Lewenstein for fruitful discussions. We
thank
the support of the EU (OLAQUI, SCALA, QICS), the National Research
Fund of Hungary OTKA (Contract No. T049234),
the Hungarian
Academy of Sciences (Bolyai Programme),
the FWF, and the Spanish MEC (Ramon y Cajal Programme,
Consolider-Ingenio 2010 project ''QOIT'').

{\it Appendix --- Proof of Observation 1.}  Fully separable states
are of the form $\rho = \sum_l p_l \rho_l^{(1)} \otimes \rho_l^{(2)}
\otimes ...\otimes \rho_l^{(N)},$ where $\sum_l p_l=1$ and $p_l>0.$
The variance, defined as $\va{A}:=\exs{A^2}-\exs{A}^2,$ is concave
in the state thus it suffices to prove that the inequalities of
Observation 1 are satisfied by pure product states. Based on the
theory of angular momentum, Eq.~(\ref{theorem1a}) is valid for all
quantum states.
For Eq.~(\ref{Jxyzineq_singlet}) one first needs that for product
states $\va{J_k}= N/4-(1/4)\sum_i \exs{\sigma_k^{(i)}}^2$ holds,
then the statement follows form the normalization of the Bloch
vector. Concerning Eq.~(\ref{Jxyzineq_spsq2}), we have to show that
$\mathfrak{Y}:=(N-1)\va{J_x}+N/2-\exs{J_y^2}-\exs{J_z^2}\geq 0.$
Using the abbreviation $x_i=\exs{\sigma_x^{(i)}},
y_i=\exs{\sigma_y^{(i)}},$ etc. this can be written as
$\mathfrak{Y}=(N-1)[N/4-(1/4)\sum_i x_i^2] - (1/4)\sum_{i \neq j}
(y_i y_j + z_i z_j) = (N-1)[N/4-(1/4)\sum_i x_i^2]- (1/4)[(\sum_{i}
y_i)^2 + (\sum_{i} z_i)^2] + (1/4)\sum_i(y_i^2+z_i^2). $ Using the
fact that $(\sum_i s_i)^2 \le N \sum_i s_i^2,$ and the normalization
of the Bloch vector, it follows that $\mathfrak{Y}\geq 0.$
Eq.~(\ref{Jxyzineq_spsq3}) can be proved in the same way. \qed


\begin{thebibliography}{99}
\bibitem{remark} Note that the condition
$[H,M_i]=0$ is  difficult to guarantee in realistic situations.
\end{thebibliography}
\end{document}